\documentclass[12pt,letterpaper]{article}
\pdfoutput=1
\usepackage{jheppub}
\usepackage{amsfonts, amsthm}
\usepackage[english]{babel}
\usepackage[utf8]{inputenc}
\usepackage{slashed}
\usepackage{mathrsfs}
\usepackage{amssymb}
\usepackage{color}
\hypersetup{unicode}

\newcommand{\eq}{\begin{equation}}
\newcommand{\feq}{\end{equation}}
\newcommand{\eqn}{\begin{eqnarray}}
\newcommand{\feqn}{\end{eqnarray}}
\font\mybb=msbm10 at 12pt
\def\bb#1{\hbox{\mybb#1}}
\def\bZ {\bb{Z}}

\def\bR {\bb{R}}

\newcommand{\ma}[1]{\mbox{$\mathcal{#1}$}}

\newcommand{\D}{{\rm d}}

\title{AdS$_5$ black strings in the stu model of FI-gauged $N=2$ supergravity}

\author{Matteo Azzola,}
\author{Dietmar Klemm,}
\author{and Marco Rabbiosi}

\affiliation{Dipartimento di Fisica, Universit\`a di Milano, and \\
INFN, Sezione di Milano, \\
Via Celoria 16, I-20133 Milano, Italy.}

\emailAdd{matteo.azzola@mi.infn.it}
\emailAdd{dietmar.klemm@mi.infn.it}
\emailAdd{marco.rabbiosi@unimi.it}
\preprint{IFUM-1056-FT}

\abstract{We analytically construct asymptotically AdS$_5$ black string solutions starting from the
four-dimensional domain wall black hole of \cite{Cacciatori:2009iz}. It is shown that its uplift gives
a black string in $d=5$ minimal gauged supergravity, with momentum along the string.
Applying instead the residual symmetries of $N=2$, $d=4$ Fayet-Iliopoulos-gauged supergravity discovered
in \cite{Cacciatori:2016xly} to the domain wall seed leads, after uplifting, to a dyonic black string that 
interpolates between AdS$_5$ and $\text{AdS}_3\times\text{H}^2$ at the horizon. A Kaluza-Klein
reduction of the latter along an angular Killing direction $\phi$ followed by a duality transformation
yields, after going back to five dimensions, a black string with both momentum along the string and
rotation along $\phi$. This is the first instance of using solution-generating techniques in gauged
supergravity to add rotation to a given seed. These solutions all have constant scalar fields. As was
shown in \cite{Cacciatori:2003kv}, the construction of supersymmetric static magnetic black strings
in the FI-gauged stu model amounts to solving the $\text{SO}(2,1)$ spinning top equations, which
descend from an inhomogeneous version of the Nahm equations.
We are able to solve these in a particular case, which leads to a generalization of the
Maldacena-Nu\~nez solution.}


\begin{document}
\maketitle
\flushbottom

\section{Introduction}

Exact solutions to Einstein's field equations and their supergravity generalizations have been playing,
and continue to play, a crucial role in many important developments in general relativity, black hole
physics, integrable systems, string theory and quantum gravity. Being highly nonlinear, coupled
partial differential equations, these are notoriously difficult to solve, sometimes even in presence
of a high degree of symmetry, for instance in supergravity where one has typically many other fields
in addition to the metric. For this reason, solution generating techniques have become a very powerful
tool to generate new solutions from a given seed. The basic idea is to reduce the action (if one has
sufficiently many commuting Killing vector fields) to three
dimensions, where all vector fields can be dualized to scalars, so that one ends up with a nonlinear
sigma model coupled to gravity. The target space symmetries can then be used to obtain new solutions
to the field equations by starting from a known one\footnote{Two of the many notable examples are
the most general rotating black hole solution of five-dimensional $N=4$ superstring vacua constructed
by Cveti\v{c} and Youm \cite{Cvetic:1996xz}, or the black Saturn found in \cite{Elvang:2007rd}.}.
In the case of four-dimensional Einstein-Maxwell
gravity, for instance, one gets a sigma model whose scalars parametrize the Bergmann space
$\text{SU}(2,1)/\text{S}(\text{U}(1,1)\times\text{U}(1))$ \cite{Breitenlohner:1987dg,Breitenlohner:1998cv}.

In gauged supergravity, supersymmetry requires a potential for the moduli (except for flat gaugings),
that generically breaks the target space symmetries. For the simple example quoted above, the addition
of a cosmological constant breaks three of the eight $\text{SU}(2,1)$ symmetries, corresponding to the 
generalized Ehlers and the two Harrison transformations. This leaves a semidirect product of a
one-dimensional Heisenberg group and a translation group $\bR^2$ as residual
symmetry \cite{Klemm:2015uba}, that cannot be used to generate new solutions.

Recently however, elaborating on earlier work \cite{Halmagyi:2013uza}, the authors
of \cite{Cacciatori:2016xly} developed a solution generating technique for $N=2$, $d=4$
Fayet-Iliopoulos (FI)-gauged supergravity as well, which essentially involves the stabilization of the 
symplectic vector of gauge couplings (FI parameters) under the action of the U-duality symmetry of the 
ungauged theory. One of the main goals of the present paper is to provide an explicit application
of the method introduced in \cite{Cacciatori:2016xly}. Namely, we start from the supersymmetric
domain wall black hole in the stu model of $N=2$, $d=4$ FI-gauged supergravity, with prepotential
$F=-X^1X^2X^3/X^0$, constructed in \cite{Cacciatori:2009iz}. We then act on it with one of the
symmetry transformations of \cite{Cacciatori:2016xly} and lift the resulting configuration to five
dimensions. This leads to a dyonic black string in minimal $N=2$, $d=5$ gauged supergravity,
with momentum along the string\footnote{Notice that the idea of relating
black strings in five-dimensional gauged supergravity to black holes in 4d gauged supergravity
goes back to \cite{Hristov:2014eza}.}.
In the next step, one performs a Kaluza-Klein
reduction along an angular Killing direction $\phi$ followed by another duality transformation.
After going back to $d=5$ one gets a black string with both momentum and rotation.
This is the first instance of using solution-generating techniques in gauged supergravity to add
rotation to a given seed.

Black strings that interpolate between $\text{AdS}_5$ at infinity and $\text{AdS}_3\times\Sigma$ near
the horizon (where $\Sigma$ denotes a two-dimensional space of constant curvature), are interesting
in their own right, since they provide a holographic realization of so-called RG flows across dimensions,
from a four-dimensional CFT to a CFT$_2$ in the IR. A first example for the gravity dual of such a scenario 
was given in \cite{Klemm:2000nj}, and subsequently Maldacena and Nu\~nez \cite{Maldacena:2000mw}
found a string theory realization in terms of D3-branes wrapping holomorphic Riemann surfaces.
Since then, many papers on this subject appeared, cf.~\cite{Kim:2005ez,Kapustin:2006hi,Donos:2008ug,
Benini:2013cda,Karndumri:2013iqa,Donos:2014eua,Benini:2015bwz,Amariti:2016mnz,Amariti:2017cyd} for an (incomplete) list of references.

The remainder of this paper is organized as follows: In the next section, we briefly introduce the
stu model of $N=2$, $d=5$ FI-gauged supergravity and the $r$-map which relates it to the corresponding
four-dimensional model. We also determine the residual duality symmetries of the latter along the
recipe of \cite{Cacciatori:2016xly}, and review the domain wall black hole seed solution constructed
in \cite{Cacciatori:2009iz}. In section \ref{sec:string-mom}, this configuration is lifted to $d=5$, and it
is shown that the result is a black string with momentum along the string. Subsequently, in
\ref{sec:dyonic-string}, we apply a duality transformation to the seed of \cite{Cacciatori:2009iz} to
get, after uplifting, a dyonic black string. This is then Kaluza-Klein reduced along an angular Killing
direction $\phi$ in sec.~\ref{sec:rot-string}, duality-transformed and lifted back to generate a black
string with both momentum and rotation. Finally, \ref{sec:run-scal} contains the construction of
supersymmetric static magnetic black strings generalizing the Maldacena-Nu\~nez solution, by
solving the $\text{SO}(2,1)$ spinning top equations. Moreover, we make some comments on a possible
inclusion of hypermultiplets. We conclude in section \ref{sec:concl} with a discussion of our results.
A summary of the supersymmetry variations of the five-dimensional
theory is relegated to an appendix.

\section{The stu model of FI-gauged $N=2$ supergravity}

The bosonic Lagrangian of $N=2$, $d=5$ FI-gauged supergravity coupled to $n_{\text v}$ vector
multiplets is given by \cite{Gunaydin:1984ak}\footnote{The indices $I,J,\ldots$ range from $1$ to
$n_{\text v}+1$, while $i,j,\ldots=1,...,n_{\text v}$.}
\begin{equation}
e^{-1}\mathscr L^{(5)} = \frac R2 - \frac12\mathcal G_{ij}\partial_{\mu}\phi^i\partial^{\mu}\phi^j -
\frac14 G_{IJ} F^I_{\mu\nu} F^{J\mu\nu} + \frac{e^{-1}}{48} C_{IJK}\epsilon^{\mu\nu\rho\sigma\lambda}
F^I_{\mu\nu} F^J_{\rho\sigma} A^K_{\lambda} - g^2 V_5\,, \label{eq:genlag}
\end{equation}
where the scalar potential reads
\begin{equation}
V_5 = V_I V_J\left(\frac92\mathcal G^{ij}\partial_i h^I\partial_j h^J - 6 h^I h^J\right)\,. \label{pot-5d}
\end{equation}
Here, $V_I$ are FI constants, $\partial_i$ denotes a partial derivative with respect to the real scalar field
$\phi^i$, and $h^I=h^I(\phi^i)$ satisfy the condition
\begin{equation}
\mathcal V\equiv\frac16 C_{IJK} h^I h^J h^K= 1\,. \label{eq:intersection}
\end{equation}
The stu model is defined by the symmetric tensor with only nontrivial component $C_{123}=1$, up to
permutations. In this case, the functions
$h^I = h^I(\phi^1,\phi^2)$ are given by
\begin{equation}
h^1 = e^{-\frac{\phi^1}{\sqrt6} - \frac{\phi^2}{\sqrt2}}\,, \qquad h^2 = e^{\frac{2\phi^1}{\sqrt6}}\,,
\qquad h^3 = e^{-\frac{\phi^1}{\sqrt6} + \frac{\phi^2}{\sqrt2}}\,, \label{eq:fun}
\end{equation}
and therefore the constraint \eqref{eq:intersection} is satisfied.

Below we will need the relationship between \eqref{eq:genlag} and four-dimensional $N=2$ FI-gauged
supergravity, with bosonic Lagrangian
\begin{equation}
e^{-1}\mathscr L^{(4)} = \frac R2 - g_{I\bar J}\partial_{\mu} z^I\partial^{\mu}\bar z^{\bar J}
+ \frac14 I_{\Lambda\Sigma} F^{\Lambda\mu\nu} F^{\Sigma}_{\mu\nu} + \frac14 R_{\Lambda\Sigma}
F^{\Lambda\mu\nu}\!\star\! F^{\Sigma}_{\mu\nu} - V_4(z,\bar z)\,, \label{eq:lag}
\end{equation}
where $R_{\Lambda\Sigma}=\text{Re}\,{\cal N}_{\Lambda\Sigma}$, $I_{\Lambda\Sigma}=\text{Im}\,
{\cal N}_{\Lambda\Sigma}$, and $\cal N$ is the period matrix that determines the couplings of the scalars
to the vector fields. $\cal N$ is determined by a homogeneous function $F$ of degree two, called the
prepotential.
The scalar potential can be written in the symplectically covariant form \cite{DallAgata:2010ejj}
\begin{equation}
\label{eq:scal_pot}
V_4 = g^{I\bar J} D_I{\mathcal L} D_{\bar J}\bar{\mathcal L} - 3{\mathcal L}\bar{\mathcal L}\,,
\end{equation}
with $\mathcal L = \langle\mathcal G,\mathcal V\rangle\equiv\mathcal G^t\Omega\mathcal V$,
where $\mathcal V$ denotes the covariantly holomorphic symplectic section,
$\mathcal G=(g^\Lambda,g_\Lambda)^t$ is the symplectic vector of FI parameters and $D$ the
K\"ahler-covariant derivative.
More details can be found
e.~g.~in \cite{Andrianopoli:1996cm,Freedman:2012zz,DallAgata:2010ejj,Cacciatori:2016xly}.

If one reduces the action \eqref{eq:genlag} to four dimensions using the
$r$-map \cite{Klemm:2016kxw}\footnote{We apologize for using the same greek indices $\mu,\nu,\ldots$
both in five and four dimensions, but the meaning should be clear from the context.}
\begin{equation}
\D s^2_5 = e^{\frac{\phi}{\sqrt3}}\D s^2_4 + e^{-\frac2{\sqrt3}\phi}(\D z + K_\mu\D x^\mu)^2\,,
\qquad A^I = B^I(\D z + K_\mu\D x^\mu) + C^I_\mu\D x^\mu\,, \label{KK5to4}
\end{equation}
\begin{equation}
\begin{split}
&z^I = B^I + i e^{-\frac{\phi}{\sqrt3}} h^I\,, \qquad e^{\cal K} = \frac18 e^{\sqrt3\phi}\,, \qquad
g V_I = \frac{g_I}{3\sqrt2}\,, \\
& g_{I\bar J} = \frac12 e^{\frac{2\phi}{\sqrt3}} G_{IJ}\,, \qquad F_{\mu\nu}^{\Lambda} =
\frac1{\sqrt2} (K_{\mu\nu}, C^I_{\mu\nu})\,, \label{dictiormap}
\end{split}
\end{equation}
one ends up with the model with cubic prepotential
\begin{equation}
F = -\frac{X^1 X^2 X^3}{X^0}\,, \label{eq:model}
\end{equation}
and FI parameters
\begin{equation}
\ma G = (0,0,0,0,0,g_1,g_2,g_3)^t\,. \label{vec-G}
\end{equation}
The theory \eqref{eq:lag} enjoys a residual symmetry that is determined by evaluating the stabilizer of
the U-duality group acting in the symplectic representation on the vector $\ma G$ of the couplings of the
theory \cite{Cacciatori:2016xly}. In the present case the embedding of $\text{SL}(2,\mathbb R)^3$ in
$\text{Sp}(8,\mathbb R)$ can be found in \cite{Cacciatori:2016xly} and the vector $\ma G$ is given by 
\eqref{vec-G}. With a slight loss of generality we impose $g_1 = g_2 = g_3\equiv g$. Then the
stabilizer algebra reads
\begin{equation}
T(a_1, a_2 ) =\left(\begin{array}{cccccccc}
0 & 0 & 0 & 0 & 0 & 0 & 0 & 0 \\
a_1 & 0 & 0 & 0 & 0 & 0 & 0 & 0 \\
a_2 & 0 & 0 & 0 & 0 & 0 & 0 & 0 \\
-a_1 -a_2 & 0 & 0 & 0 & 0 & 0 & 0 & 0 \\ 
0 & 0 & 0 & 0 & 0 & -a_1  & -a_2 & a_1 + a_2 \\
0 & 0 & a_1 + a_2 & - a_2 & 0 & 0 & 0 & 0 \\
0 & a_1 + a_2 & 0 & -a_1 & 0 & 0 & 0 & 0 \\
0 & -a_2 & -a_1 & 0 & 0 & 0 & 0 & 0
\end{array}\right)\,, \label{talpha}
\end{equation}
a two-dimensional abelian nilpotent subalgebra of order three of $\text{sp}(8,\mathbb R)$.

The solution generating technique of \cite{Cacciatori:2016xly} consists in the transformation of the seed 
configuration $(\ma V,\ma G,\ma Q)$ $\mapsto$ $(S\ma V,\ma G,S\ma Q)$, where $S$ is an element of the 
stabilizer group $\ma S_{\ma G}$. Here, $\ma Q=(p^\Lambda, q_\Lambda)^t$ is the symplectic vector
of magnetic and electric charges that enters the field strengths as follows:
For a static solution of the type that we shall consider below,
\begin{equation}
\D s^2 = -e^{2U}\D t^2 + e^{-2U}\left(\D Y^2 + e^{2\psi}(\D\theta^2 + \sinh^2\!\theta\D\phi^2)\right)\,,
\label{eq:seed2}
\end{equation}
the Maxwell equations are solved by
\begin{equation}
F^\Lambda_{tY} = \frac12 e^{2(U - \psi)} I^{\Lambda\Sigma} (R_{\Sigma\Gamma} p^\Gamma - q_\Sigma)\,,
\qquad F^\Lambda_{\theta\phi} = \frac{p^\Lambda}2\sinh\theta\,,
\end{equation}
and the corresponding dual field strengths
\begin{equation}
F_{\Lambda\mu\nu} = R_{\Lambda\Sigma}F^{\Sigma}_{\mu\nu} - \frac12 I_{\Lambda\Sigma}
e\epsilon_{\mu\nu\rho\sigma} F^{\Sigma\rho\sigma}
\end{equation}
are
\begin{displaymath}
F_{\Lambda tY} = \frac12 e^{2(U - \psi)}\left(I_{\Lambda\Sigma} p^\Sigma + R_{\Lambda\Gamma}
I^{\Gamma\Omega} R_{\Omega\Sigma} p^\Sigma - R_{\Lambda\Gamma} I^{\Gamma\Omega}
q_\Omega\right)\,, \qquad F_{\Lambda\theta\phi} = \frac{q_\Lambda}2\sinh\theta\,.
\end{displaymath}
The charge vector $\ma Q$ is therefore completely determined by the $(\theta\phi)$-components
of $F^\Lambda$ and $F_\Lambda$.

In \cite{Cacciatori:2009iz}, various BPS solutions to the theory \eqref{eq:lag} were
constructed, by using the recipe of \cite{Cacciatori:2008ek}, where all timelike supersymmetric
backgrounds of $N=2$, $d=4$ FI-gauged supergravity are classified\footnote{For a classification of the
null case cf.~\cite{Klemm:2009uw}.}.
In particular, for the prepotential
$F=-X^1 X^2 X^3/X^0$, the solution reads\footnote{To translate between \eqref{eq:lag} and the
conventions of \cite{Cacciatori:2009iz} take $g_{\text{CK}}\rightarrow g/2$ and
$G_{\text{CK}}\rightarrow\frac1{8\pi}$.}
\begin{equation}
\D s^2 = -4 b^2\D t^2 + \frac1{b^2}\frac{y\D y^2}{cy + 2gp} + \frac{y^3}{b^2}(\D\theta^2 +
\sinh^2\!\theta\D\phi^2)\,, \label{eq:met-domainwall}
\end{equation}
where
\begin{equation}
b^4 = \frac{8g_1 g_2 g_3 y^{\frac92}}{H^0(cy + 2gp)^{\frac32} }\,, \qquad
H^0 = \frac{2q_0}{3g^2 p^2 y^{\frac32}}(cy + 2gp)^{\frac12}(cy - gp) + h^0\,,
\end{equation}
with field strengths and scalars
\begin{eqnarray}
F^0 &=& 4\D t\wedge\D (H^0)^{-1}\,, \qquad F^I = \frac{p^I}2\sinh\theta\D\theta\wedge\D\phi\,,
\nonumber \\
z^I &=& i\tau^I = i\frac{\sqrt{g_1 g_2 g_3}}{\sqrt2 g_I}\frac{\sqrt{H^0} y^{\frac34}}{(cy + 2gp)^{\frac14}}\,,
\label{eq:fields}
\end{eqnarray}
and the magnetic charges are constrained by $g_I p^I\equiv gp=2/3$ (no summation over
$I$)\footnote{This follows from eq.~(2.23) of \cite{Cacciatori:2009iz} with $\kappa=-1$, by taking
into account that $p^I_{\text{here}}=8\pi p^I_{\text{CK}}$.} for
$I=1,2,3$. This field configuration preserves two real supercharges, while the near-horizon limit is a
half-BPS attractor point $\text{AdS}_2\times\text{H}^2$.
The range of the parameters is $q_0<0$, $p^I>0$, $c>0$, $h^0>3|q_0|c^{3/2}/2$ and $g_I>0$.
\eqref{eq:met-domainwall} has a horizon in $y=0$ and asymptotes to a curved domain wall (whose
worldvolume is an open Einstein static universe) for $y\to\infty$.

The $(\theta,\phi)$-components of the dual field strengths are given by
\begin{equation}
F_{0\theta\phi} = \frac{q_0}2\,, \qquad F_{I\theta\phi} = 0\,.
\end{equation}
From this, together with \eqref{eq:fields}, one can easily read off the charge vector $\ma Q$.

\section{Black string with momentum along $\partial_z$}
\label{sec:string-mom}

The $r$-map \eqref{KK5to4}, \eqref{dictiormap} can be used to uplift the field configuration 
\eqref{eq:met-domainwall}, \eqref{eq:fields} to five dimensions. If we define $y=:r^2$, the resulting
metric in $d=5$ reads
\begin{equation}
\D s^2 = 2\left(\frac1{H^0 b}\right)^{\frac23}\left(\frac1{b^2}\frac{4 r^4\D r^2}{c r^2 + 2 g p} +
\frac{r^6}{b^2}(\D\theta^2 + \sinh^2\!\theta\D\phi^2)\right) + \frac14 (H^0 b)^{\frac43}
\left(\D z^2 - \frac{8\sqrt2}{H^0}\D t\D z\right)\,, \label{eq:metrbs}
\end{equation}
with
\begin{equation}
b^4 = \frac{8 g_1 g_2 g_3 r^9}{H^0  (c r^2 + 2 g p)^{\frac32} }\,, \qquad
H^0 = \frac{2 q_0}{3 g^2 p^2 r^{3}} (c r^2 + 2 g p)^{\frac12} (c r^2 - g p) + h^0\,. \label{H^0(r)}
\end{equation}
The fluxes and the scalars are given by
\begin{equation}
F^I = \frac{p^I}{\sqrt 2}\sinh\theta\D\theta\wedge\D\phi\,, \qquad h^I =
\frac{(g_1 g_2 g_3)^{1/3}}{g_I}\,. \label{eq:flsca}
\end{equation}
The configuration \eqref{eq:metrbs}, \eqref{eq:flsca} satisfies the BPS equations \eqref{eq:BPS} with
Killing spinor
\begin{equation}
\epsilon = Y(r)(1 + i\Gamma^{32})(1 - \Gamma^1)\epsilon_0\,,
\end{equation}
where $\epsilon_0$ is a generic constant Dirac spinor. Since $\epsilon_0$ is subject to a double
projection, the solution is one quarter BPS. It has a horizon in $r=0$, where the spacetime becomes
$\text{AdS}_3\times\text{H}^2$, while asymptotically it approaches what is commonly termed
magnetic \cite{Hristov:2011ye} $\text{AdS}_5$.

The solution \eqref{eq:metrbs}, \eqref{eq:flsca} was first constructed in \cite{Bernamonti:2007bu}, by
using the results of \cite{Gauntlett:2003fk}, where all supersymmetric backgrounds of minimal
gauged supergravity in five dimensions were classified. It describes a gravitational wave propagating
along a magnetic black string, which can be seen by introducing the new coordinates
\begin{equation}
\rho = \frac{\ell^2 (H^0)^{1/3} b^{4/3}}{\sqrt2 r^3}\,, \qquad u = \frac{8^{1/4} z}{\ell^{1/2} c^{3/4}}\,,
\qquad v = -\frac{32 t}{\ell^{1/2} c^{3/4} 8^{1/4}}\,,
\end{equation}
where $\ell$ is defined by
\eq
\frac2{\ell^2} = (g_1 g_2 g_3)^{2/3}\,,
\feq
such that the cosmological constant is $\Lambda=-6\ell^{-2}$. The metric \eqref{eq:metrbs} becomes
\eq
\D s^2 = \frac{\ell^2\D\rho^2}{\rho^2 h^2} + \frac{\ell^4}{\rho^2}(\D\theta^2 + \sinh^2\!\theta
\D\phi^2) + \frac{\ell^2}{\rho^2}h^{3/2}\left(H^0\D u^2 + \D u\D v\right)\,, \label{grav-wave}
\feq
with
\eq
h = 1 - \frac{\rho^2}{3\ell^2}\,.
\feq
In the new radial coordinate $\rho$, the wave profile reads\footnote{To get \eqref{grav-wave} from (4.21)
of \cite{Bernamonti:2007bu}, set $k=-1$ there (which implies ${\cal P}=0$) and ${\cal H}_2=1$. Then
the Heun equation (4.25) of \cite{Bernamonti:2007bu} boils down to a simple differential equation
that is solved by \eqref{wave-profile}.}
\eq
H^0 = \frac{3q_0 c^{3/2}}{2h^{3/2}}\left(1 - \frac{\rho^2}{2\ell^2}\right) + h^0\,. \label{wave-profile}
\feq
In the near-horizon limit $\rho\to\sqrt3\ell$, \eqref{grav-wave} approaches
\eq
\D s^2 = (\D u + l\hat\rho\D v)^2 + l^2\left(\frac{\D\hat\rho^2}{4\hat\rho^2} - \hat\rho^2\D v^2
\right) + \frac{\ell^2}3(\D\theta^2 + \sinh^2\!\theta\D\phi^2)\,, \label{grav-wave-nh}
\feq
where we eliminated the constant $h^0$ by shifting $v\to v-h^0 u$, and subsequently rescaled
\eq
u \to\frac{2u}{|q_0|^{1/2}c^{3/4}}\,, \qquad v\to (|q_0|/2)^{1/2} (3c)^{3/4}\ell v\,.
\feq
Moreover we defined
\eq
\hat\rho = \left(\sqrt3 - \frac{\rho}{\ell}\right)^{3/2}\,, \qquad l = \frac{2\ell}3\,.
\feq
Note that the AdS$_3$ factor in \eqref{grav-wave-nh} is written as a Hopf-like fibration over AdS$_2$,
in which $\partial_v$ is a null direction.

The momentum $L_z$ along the string can be computed as a Komar integral associated to the
Killing vector $\xi=\partial_u$,
\begin{equation}
L_z = \frac1{16\pi G_5}\oint_{\partial V} dS^{\mu\nu}\nabla_\mu\xi_\nu\,,
\end{equation}
where $\partial V$ is the boundary of a spacelike hypersurface $V$ of constant $v$. $\partial V$ is
defined by $\rho=\text{const.}\to 0$. The oriented measure on $\partial V$ reads
\eq
dS^{\mu\nu} = (\zeta^\mu n^\nu - n^\mu\zeta^\nu)\sqrt\sigma\D\theta\D\phi\D u\,,
\feq
where
\eq
n = \frac{\rho}{\ell h^{3/4}{H^0}^{1/2}}(\partial_u - 2H^0\partial_v)
\feq
denotes the unit normal to $V$, while $\zeta=-\rho h\ell^{-1}\partial_\rho$ and $\sigma$ is the
determinant of the induced metric on $\partial V$,
\eq
\sqrt\sigma = \frac{\ell^5}{\rho^3}h^{3/4}{H^0}^{1/2}\sinh\theta\,.
\feq
This leads to
\eq
L_z = \frac{L\ell q_0 c^{3/2}(\mathfrak{g} - 1)}{16 G_5}\,.
\feq
Here $\mathfrak{g}=2,3,\ldots$ denotes the genus of the Riemann surface to which $\text{H}^2$
is compactified, and $L$ is the period of $u$ in \eqref{grav-wave-nh}.

The Bekenstein-Hawking entropy of the solution \eqref{grav-wave} is given by\footnote{To compute the
horizon area, we took a section of constant $\hat\rho$ and $v$ in \eqref{grav-wave-nh}.}
\eq
S_{\text{BH}} = \frac{A_{\text{hor}}}{4G_5} = \frac{\ell^2 L\pi(\mathfrak{g} - 1)}{3G_5}\,.
\feq
It would be very interesting to generalize \eqref{grav-wave}, for instance by allowing for a nontrivial
dependence of the wave profile on the `angular' coordinates $\theta,\phi$. According to the governing 
equations for the wave profile (cf.~[4.22] and [4.26] of \cite{Bernamonti:2007bu}) this is possible,
and a Kaluza-Klein reduction of such a solution to four dimensions would lead to a static black hole
with dipole- or higher multipole charges. Work in this direction is in progress.

\section{Dyonic black string with momentum}
\label{sec:dyonic-string}

As we said, the results of \cite{Cacciatori:2016xly} can be used to generate a new configuration from the
seed solution \eqref{eq:met-domainwall} and the respective fluxes and scalars \eqref{eq:fields}.
It would be interesting to fix the 
parameters of the symmetry transformation $T(a_1, a_2)$ in \eqref{talpha} to get a vanishing
graviphoton and therefore a static metric in five dimensions.
However, as the following calculation shows, this results to be impossible.
Starting form the fluxes\footnote{Note that $q_0=-|q_0|$. Moreover, if we choose $g_1=g_2=g_3\equiv g$,
the constraint $g_Ip^I=gp$ satisfied by the seed implies $p^1=p^2=p^3=p$.} 
\begin{equation}
\ma Q = (0, p, p, p, -|q_0|, 0, 0, 0)^t\,,
\end{equation}
the action of $U=e^{T(a_1,a_2)}$, with $T(a_1,a_2)$ given by \eqref{talpha}, generates
\begin{equation}
\ma Q^{\prime} =\left(\begin{array}{c}
0 \\
p \\
p \\
p \\
-|q_0| + (a_1 a_2 + a_1^2 + a_2^2 )p \\
-a_1 p \\
-a_2 p \\
(a_1 + a_2 ) p \\
\end{array}\right)\,.
\label{eq:Qp}
\end{equation}
The scalars acquire a constant axion,
\begin{equation}
z^{1\prime} = z^1 + a_1\,, \qquad z^{2\prime} = z^2 + a_2\,, \qquad z^{3\prime} = z^3 - a_1 - a_2\,.
\label{eq:scalar}
\end{equation}
Looking at \eqref{eq:Qp} one may think that the graviphoton can be set to zero by choosing properly the 
parameters $a_1,a_2$. This is however not the case, as can be seen from the field strengths in presence of
axions, that read
\begin{equation}
F^\Lambda = \frac{S^\Lambda b^2}{y^{5/2}(cy + 2gp)^{1/2}}\D t\wedge\D y + 
\frac{p^\Lambda}2\sinh\theta\D\theta\wedge\D\phi\,,
\end{equation}
with
\begin{equation}
S^\Lambda =\frac{|q_0|}{\tau^1\tau^2\tau^3}(1, a_1, a_2, -a_1 - a_2)^t\,, \qquad p^\Lambda = 
(0, p, p, p)^t\,.
\end{equation}
The uplifted metric is still \eqref{eq:metrbs}.
The field strengths and scalars in five dimensions become respectively
\begin{equation}
F^I = \frac{2\sqrt2 b^2 S^I}{r^4(c r^2 + 2gp)^{1/2}}\D t\wedge\D r + \frac{p^I}{\sqrt2}\sinh\theta
\D\theta\wedge\D\phi\,, \qquad h^I = 1\,,
\label{eq:fsrbs}
\end{equation}
where now
\begin{equation}
S^I = \frac{|q_0|}{\tau^1\tau^2\tau^3}(a_1, a_2, -a_1 -a_2)^t\,, \qquad p^I = (p,p,p)^t\,, \label{eq:veccha}
\end{equation}
so we have generated two additional electric charge parameters. Note that the metric remains untouched
by the duality rotation. This solution describes again a flow between magnetic AdS$_5$ and 
$\text{AdS}_3\times\text{H}^2$, and preserves the same amount of supersymmetry as before, as can
be easily seen by using the Killing spinor equations following from \eqref{eq:BPS}.

\section{Dyonic black string with both momentum and rotation}
\label{sec:rot-string}

We now want to generate a rotating black string by applying the same technique as above.
The starting point is the seed metric \eqref{eq:metrbs}, with fluxes
and scalars \eqref{eq:fsrbs}. After a Kaluza-Klein reduction to four dimensions along the angular Killing 
direction $\partial_\phi$, one gets
\begin{eqnarray}
\D s^2 &=&\sinh\theta\left(\frac{\sqrt2(c r^2 + 2gp)^{1/2}}{g^3 r^2}\D r^2 + \frac{(c r^2 + 2gp)^{3/2}}
{2\sqrt2 g^3}\D\theta^2 + 2 r^3\left(\frac{H^0}{4\sqrt2}\D z  - \D t\right)\!\D z\right)\,, \nonumber \\
A^\Lambda &=&\left(0, \frac{4 q_I}{H^0}\D t\right)\,, \qquad z^I = \frac{p^I}{\sqrt2}\cosh\theta
+ i\frac{(c r^2 + 2gp)^{1/2}\sinh\theta}{\sqrt2 g_I}\,, \label{eq:metrbs1}
\end{eqnarray}
where $q_3=-q_1-q_2$. Applying the duality transformation \eqref{talpha} leads to a configuration
with a nonvanishing graviphoton,
\begin{equation}
\ma Q = (0,0,0,0,0,q_1,q_2,-q_1-q_2)\mapsto\ma Q^\prime = (0,0,0,0,\omega,q_1,q_2,-q_1-q_2)^t\,,
\end{equation}
where $\omega = -a_1 (2 q_1 +q_2) - a_2 (q_1 + 2 q_2)$, and the scalar fields acquire a real constant
part as in \eqref{eq:scalar}. Lifting the solution back to $d=5$ gives
\begin{equation}
\begin{split}
\D s^2 =\frac{2\D r^2}{g^2 r^2} &+ \frac{cr^2 + 2gp}{2g^2}\D\Omega^2_{\text{H}^2} +
\frac{2\sqrt2 g r^3}{(cr^2 + 2gp)^{1/2}}\left(\frac{H^0}{4\sqrt2}\D z - \D t\right)\!\D z \\
&+ 4\sqrt2\omega\frac{cr^2 + 2gp}{g^2 H^0}\sinh^2\!\theta\left(\D\phi +
\frac{2\sqrt2\omega}{H^0}\D t\right)\!\D t\,,
\end{split}
\label{eq:metrbs2}
\end{equation}
\begin{equation}
A^I = \frac{p^I}{\sqrt2}\cosh\theta\D\phi + \frac4{H^0}\left(q_I + \omega s^I + \frac{\omega p^I}{\sqrt2}
\cosh\theta\right)\D t\,, \qquad h^I =1\,, \label{A^I-rot}
\end{equation}
with $s^I = (a_1, a_2 , -a_1-a_2)$ and $H^0$ was defined in \eqref{H^0(r)}.

The near-horizon limit $r\to 0$ of \eqref{eq:metrbs2} leads to the metric
\begin{equation}
\D s^2 = \frac{|q_0|}{3p}(\D z - \hat\rho\D t)^2 + \frac{l^2\D\hat\rho^2}{4\hat\rho^2} - \frac{|q_0|}{3p}
\hat\rho^2\D t^2 + \frac pg\left[\D\theta^2 + \sinh^2\!\theta(\D\phi + 2\hat\rho\omega\D t)^2\right]\,,
\label{nh-rot-string}
\end{equation}
where 
\begin{equation}
\hat\rho\equiv\frac{3\sqrt{gp}}{|q_0|}r^3\,.
\end{equation}
\eqref{nh-rot-string} represents a deformation of \eqref{grav-wave-nh}, i.e., of 
$\text{AdS}_3\times\text{H}^2$.

For $r\to\infty$ \eqref{eq:metrbs2} approaches again magnetic AdS$_5$, as can be easily shown by
using some simple coordinate transformations.

It would be interesting to check whether the solution \eqref{eq:metrbs2}, \eqref{A^I-rot} is still BPS,
or, more generally, if the solution-generating technique based on \cite{Cacciatori:2016xly} preserves
supersymmetry.

\section{Solutions with running scalars}
\label{sec:run-scal}

In the appendix of \cite{Cacciatori:2003kv} the problem of finding one quarter magnetic BPS strings
with running scalars is reduced to solve a system of three first-order differential equations. The metric
is given by\footnote{In this section we choose $V_I=\frac13$ in \eqref{pot-5d} without loss
of generality.}
\begin{equation}
\D s^2 = e^{2V}(-\D t^2 +\D z^2) + e^{2W}(\D u^2 +\D \Omega_\kappa^2)\,, \label{metr-string-CKS}
\end{equation}
with
\eq
\D\Omega_\kappa^2 = \D\theta^2 + \frac{\sin^2\!\sqrt\kappa\theta}\kappa\D\phi^2\,,
\qquad\kappa = 0,\pm 1\,,
\feq
\eq
e^{2V} = (x^1 x^2 x^3)^{-\frac13} e^{-g\int(x^1 +x^2 +x^3)\D u}\,, \qquad
e^{2W} = (x^1 x^2 x^3)^{\frac23}\,,
\feq
and the $x^I(u)$ define the scalar fields according to $h^I=x^I/(x^1 x^2 x^3)^{1/3}$. They are
determined by the system\footnote{In the following we set $g=1$.}
\begin{equation}
\begin{split}
&y^{1\prime}= y^2 y^3 + Q^1\,, \\
&y^{2\prime}= y^1 y^3 + Q^2\,, \\
&y^{3\prime}= y^1 y^2 + Q^3\,, \label{eq:sys}
\end{split}
\end{equation}
where a prime denotes a derivative w.r.t.~the radial coordinate $u$ and
\begin{equation}
\begin{split}
&y^1= x^1 + x^2 - x^3\,, \\
&y^2= x^1 - x^2 - x^3\,, \\
&y^3= x^1 - x^2 + x^3\,.
\end{split}
\end{equation}
The fluxes are purely magnetic, i.e., $F^I_{\theta\phi}= \sqrt\kappa q^I\sin\sqrt\kappa\theta$,
and the parameters $Q^I$ are defined by
\begin{equation}
\begin{split}
&Q^1=-\kappa ( q^1 + q^2 - q^3)\,, \\
&Q^2= -\kappa ( q^1 - q^2 - q^3)\,, \\
&Q^3=-\kappa ( q^1 - q^2 + q^3)\,.
\end{split}
\end{equation}
\eqref{eq:sys} can be derived from an inhomogeneous version of the $\text{SU(2)}$ Nahm
equations \cite{Nahm:1982jt,Hitchin:1983ay,Donaldson:1984tm}
\eq
\frac{\D T^I}{\D u} = \epsilon^{IJK}\left[T^J, T^K\right] + S^I
\feq
(where the $T^I$ and $S^I$ take values in the Lie algebra $\text{su}(2)$) by setting $T^I=y^I\sigma^I$,
$S^I=Q^I\sigma^I$ (no summation over $I$), and the $\sigma^I$ denote Pauli matrices. Notice that
for $Q^I=0$ this leads to the Ercolani-Sinha solution \cite{Ercolani:1989tp}, which is given in terms of
elliptic functions. \eqref{eq:sys} can be written as
\eq
y^{I\prime} = {C^I}_{JK} y^J y^K + Q^I\,, \label{spinn-top}
\feq
and thus its symmetries are determined by the transformations that leave the tensor ${C^I}_{JK}$
invariant, $T^{-1}CTT=C$. Unfortunately this implies $T=1$. The discrete symmetry group of
\eqref{spinn-top}, which is easily shown to be $(\bZ_2)^6\times\bZ_3$, is not very useful for
generating new solutions from known ones.

The system \eqref{spinn-top} is equivalent to the $\text{SO}(2,1)$ spinning top equations, which are
given by
\begin{equation}
\begin{split}
&I_1\omega_1^\prime = (I_2 - I_3)\omega_2\omega_3 + M_1\,, \\
&I_2\omega_2^\prime = -(I_3 - I_1)\omega_3\omega_1 + M_2\,, \\
&I_3\omega_3^\prime = (I_1 - I_2)\omega_1\omega_2 + M_3\,, \label{spinn-top-SO(2,1)}
\end{split}
\end{equation}
where $I_K$ are the principal moments of inertia and $M_K$ represents an external torque. If we set
\eq
\omega_1 = \lambda_1 y^1\,, \qquad \omega_2 = \lambda_2 y^2\,, \qquad \omega_3 = \lambda_3 y^3\,,
\feq
where
\begin{displaymath}
\lambda_1^2 = \frac{-I_2 I_3}{(I_3 - I_1)(I_1 - I_2)}\,, \qquad
\lambda_2^2 = \frac{I_3 I_1}{(I_1 - I_2)(I_2 - I_3)}\,, \qquad
\lambda_3^2 = \frac{-I_1 I_2}{(I_2 - I_3)(I_3 - I_1)}\,,
\end{displaymath}
\eqref{spinn-top-SO(2,1)} reduces to \eqref{spinn-top}. Here we assumed (without loss of generality)
$I_1>I_2>I_3>0$. Then all $\lambda_K^2$ are positive. Note that, as compared to Euler's equations,
there is a minus sign on the rhs of the second of \eqref{spinn-top-SO(2,1)}, which is the reason for
the terminology $\text{SO}(2,1)$. In fact, in the untorqued case $M_K=0$, the
eqns.~\eqref{spinn-top-SO(2,1)} can be derived from the Hamiltonian
\eq
H = \frac{\ell_1^2}{2I_1} - \frac{\ell_2^2}{2I_2} + \frac{\ell_3^2}{2I_3}\,, \label{H-spinn-top}
\feq
by using the Poisson brackets
\eq
[\ell_I, \ell_J] = {\epsilon_{IJ}}^K\ell_K\,, \label{PB}
\feq
as well as $\ell_I^\prime=[\ell_I,H]$. In \eqref{PB}, $\epsilon_{123}=1$ and the indices of the
Levi-Civita tensor are raised with the Minkowski metric $\eta=\text{diag}(1,-1,1)$. W.r.t.~the Euler top,
\eqref{H-spinn-top} has one kinetic term that comes with the `wrong' sign.

In \cite{Cacciatori:2003kv} a particular solution to \eqref{eq:sys} is found. We will now show a different
way to integrate these equations. Imposing $Q^1=Q^2=0$, i.e., $q^2 =0$ and $q^1 = q^3$, and
defining $y_\pm= y^1\pm y^2$ one finds that
\begin{equation}
y_\pm = k_\pm e^{\int y^3\D u}\,, \qquad
y^{3\prime} = \frac14\left(k_{+}^2 e^{2\int y^3\D u} - k_{-}^2 e^{-2\int y^3\D u}\right) + Q^3\,,
\end{equation} 
where $Q^3=-2\kappa q^1$ and $k_\pm$ are integration constants. The Dirac-type charge
quantization condition $3gV_Iq^I=1$ of \cite{Cacciatori:2003kv} implies $2gq^1=1$,
and thus $Q^3=-\kappa$. Introducing a new radial
coordinate $y=-\int y^3\D u$, the last equation becomes
\begin{equation}
y^{\prime\prime}= \frac14\left(k_-^2 e^{2y} - k_+^2 e^{-2y}\right) + \kappa\,,
\end{equation}
which can be integrated once to give\footnote{The plus sign corresponds to an unphysical solution
with negative scalars $h^I$. A possible additive integration constant can be eliminated by shifting $y$.} 
\begin{equation}
y^{\prime} = \frac{\D y}{\D u} = -y^3 = -\sqrt{\frac14\left(k_-^2 e^{2y} + k_+^2 e^{-2y}\right) +
2\kappa y}\,.
\end{equation}
This leads to the metric
\begin{equation}
\D s^2 = (x^1 x^2 x^3)^{-\frac13} e^{\int\frac{x^1 +x^2 +x^3}{y^3}\D y}(-\D t^2 + \D z^2) +
(x^1 x^2 x^3)^{\frac23}\left(\frac{\D y^2}{(y^3)^2} + \D\Omega_{\kappa}^2\right)\,,
\label{eq:met}
\end{equation}
where
\begin{equation}
\begin{split}
&x^1 = \frac14\left(k_+ e^{-y} + k_- e^y + \sqrt{k_+^2 e^{-2y} + k_-^2 e^{2y} +
8\kappa y}\right)\,, \\
&x^2 = \frac{k_-}2 e^y\,, \\
&x^3 = \frac14\left(-k_+ e^{-y} + k_- e^y + \sqrt{k_+^2 e^{-2y} + k_-^2 e^{2y} +
8\kappa y}\right)\,. \label{eq:scalars}
\end{split}
\end{equation}
In what follows we assume $k_->0$. Then asymptotically for $y\to\infty$ the geometry becomes
(magnetic) AdS$_5$,
\begin{equation}
\D s^2 = \frac{2 e^{2y}}{k_-}(-\D t^2 + \D z^2) + \D y^2 + \frac{k_-^2 e^{2y}}4\D\Omega_{\kappa}^2\,.
\end{equation}
For generic integration constants $k_\pm$ the metric becomes singular at a certain point and the solution 
does not have a horizon. However, in the case $\kappa=-1$, if $k_\pm$ are related by\footnote{The
other possibility $ k_\pm = e^{\pm a}\sqrt{4 a\mp 2}$ is related to \eqref{eq:cons} by
the $\mathbb Z_2$ symmetry $x^1\leftrightarrow x^3$ and corresponds to negative $h^I$.}
\begin{equation}
k_- =  e^{-a}\sqrt{4 a + 2}\,, \qquad k_+ = - e^a\sqrt{4 a - 2}\,,
\label{eq:cons}  
\end{equation}
where $a$ denotes an arbitrary parameter, there is a horizon for $y=a$, where the solution approaches
$\text{AdS}_3\times\text{H}^2$, as we will show below. In the cases $\kappa=0,1$ a similar reasoning
cannot be done, and the metric has no event horizon.

\eqref{eq:cons} are real for $a\geq1/2$. We note that, if \eqref{eq:cons}
holds, one has
\begin{equation}
y^3(y) = \sqrt{2a\cosh(2(a - y)) - \sinh(2(a - y)) - 2y}\,,
\end{equation}
and thus
\begin{equation}
\frac{\D(y^3)^2}{\D y} = 2\cosh(2(a-y)) - 4a\sinh(2(a - y)) - 2\geq 2(e^{2(y - a)} -1)\geq 0\,.
\end{equation}
$y^3$ is always well-defined for $y\geq a$ and becomes zero at the horizon, $y^3(a)=0$. The scalar
fields \eqref{eq:scalars} are positive in the whole range $a\le y<\infty$.

The value $a=1/2$ is special since it corresponds to the limit in which $x^1=x^3$, i.e., $\phi^2 =0$.
This truncation leads to the solution of \cite{Maldacena:2000mw} with two nonzero and equal fluxes.
Indeed, one easily verifies that $\phi^1$ and the function $W$ appearing in the
metric \eqref{metr-string-CKS} satisfy the equation
\begin{equation}
e^{2W + \frac{\phi^1}{\sqrt6}} = e^{2W - \frac{2\phi^1}{\sqrt 6}} + \frac{\sqrt6 W + 2\phi^1}{2\sqrt6}
+ \frac14\,,
\end{equation}
which is precisely equ.~(17) of \cite{Maldacena:2000mw}. The black strings defined
by \eqref{eq:met}, \eqref{eq:scalars} represent thus generalizations of the Maldacena-Nu\~nez
solution. Note that also the latter was not known analytically up to now.

The near-horizon limit $y\to a$ can be obtained from the expansion
\begin{equation}
\begin{split}
&\frac{x^1 + x^2 + x^3}{y^3} = \sqrt{\frac{1 + 2a}{2a}}\frac1{y - a} + O((y - a)^0)\,, \\
&\frac{(y^3)^3}{x^1 x^2 x^3} = 32a\sqrt{\frac{2a}{1 + 2a}}(y - a)^3 + O((y - a)^4)\,, \\
& x^1 x^2 x^3 = \sqrt{\frac{1 + 2a}{32}} +  O((y - a))\,.
\end{split}
\end{equation}
Introducing the new radial coordinate 
\begin{equation} 
{\hat u}^2 = (y - a)^{-\sqrt{\frac{1 + 2a}{2a}}}\,,
\end{equation}
the metric \eqref{eq:met} becomes for ${\hat u}\to 0$  
\begin{equation}
\D s^2 = \frac1{\hat u^2}\left[-\D t^2 + \D z^2 + \frac{\D\hat u^2}{(2 + 4a)^{2/3}}\right]
+ \left(\frac{1 + 2a}{32}\right)^{1/3}\!\D\Omega_{-1}^2\,,
\end{equation}
which is $\text{AdS}_3\times\text{H}^2$.

The central charge of the two-dimensional SCFT dual to the near-horizon configuration is given
by \cite{Maldacena:2000mw,Klemm:2016kxw}
\begin{equation}
c = \frac{3 R_{\text{AdS}_3}}{2 G_3} = \frac{6\pi(\mathfrak{g} - 1)R_{\text{AdS}_3}
R_{\Sigma_{\mathfrak{g}}}^2}{G_5}\,,
\end{equation}
where $\mathfrak{g}=2,3,\ldots$ is the genus of the Riemann surface $\Sigma_{\mathfrak{g}}$ to which
$\text{H}^2$ is compactified. The values of the curvature radii are
\begin{equation}
R_{\text{AdS}_3} = \frac1{(2 + 4a)^{1/3}}\,, \qquad R_{\Sigma_{\mathfrak{g}}}^2 =
\frac{(1 + 2a)^{1/3}}{2^{5/3}}\,, \label{curv-radii}
\end{equation}
with
\begin{equation}
2a = \sqrt{1+ \left(\frac{k_+ k_-}2\right)^2}\,.
\end{equation}
For the truncation $\phi^2 =0$, which means $k_+=0$, one has $R_{\text{AdS}_3}=2^{-2/3}$,
i.e.~the value found in \cite{Maldacena:2000mw}.

Using \eqref{curv-radii}, the central charge can be written in the form
\begin{equation}
c = \frac{6\pi(\mathfrak{g} - 1)}{4 G_5} = 3N^2(\mathfrak{g} - 1)\,,
\end{equation}
where we used the AdS/CFT dictionary $N^2=\pi/(2G_5 g^3)$ (with $g=1$).
Near the conformal boundary the scalar fields behave like
\begin{equation}
\frac{2\phi^1}{\sqrt6}\sim 2Q^3 y e^{-2y}\,, \qquad \sqrt2\phi^2\sim -\frac{k_+}{k_-} e^{-2y}\,,
\end{equation}
and thus are read in the dual SCFT as an insertion and an expectation value of an operator of scaling 
dimension $\Delta=2$.
The relevant deformation of the dual superpotential relative to $\phi^1$ is described
in \cite{Maldacena:2000mw}, while $\phi^2$ is a marginal deformation of two-dimensional
$N=(4,4)$ SYM theory. Thus, the solution does not describe the gravity dual of 2d 
$N=(2,2)^*$ SYM \cite{Nian:2017usa}.
The constant $a$ represents the physical scale of the energy in the renormalization group flow at which
the IR fixed point appears, but which being a CFT is independent of the energy scale.

\subsection{Inclusion of hypermultiplets}

It is worthwhile to note that with running hyperscalars the BPS equations, (5.17) of \cite{Klemm:2016kxw}, can be simplified to a system for which the number of equations equals the number of the scalar fields
in the model. The idea is basically the same as that of \cite{Cacciatori:2003kv}:
Introducing a new radial coordinate $R$ by $\D R=e^{-\psi}\D r$ and the rescaled scalars
$y^I=e^{\psi-2T}h^I$, where $\psi(r)$ and $T(r)$ are metric functions defined in equ.~(3.1) of
\cite{Klemm:2016kxw} and $r$ denotes the radial coordinate used there, the system (5.17) of
\cite{Klemm:2016kxw} boils down to
\begin{equation}
\begin{split}
&\psi = \int 9\kappa g^2\ma L_y \D R\,, \qquad e^{3\psi - 6T} = \frac16 C_{IJK} y^I y^J y^K\,, \\
&y^{I\prime} - 9 g^2\kappa(\ma L_y y^I - \ma Q^x P^x_J G^{IJ}_y) - p^I = 0\,, \\
&q^{u\prime} = -\frac92\kappa g^2 h^{uv}\partial_v\ma L_y\,,
\end{split}
\end{equation} 
where
\begin{displaymath}
\ma L_y = \ma Q^x P^x_I y^I\,, \qquad G^{IJ}_y = -C^{IJK} C_{KLM} y^L y^M + 2 y^I y^J\,, \qquad
C^{IJK}\equiv\delta^{IL}\delta^{JM}\delta^{KN} C_{LMN}\,.
\end{displaymath}
Even if the complete integration of these equations in a particular model remains a hard task, this partial integration can be considered as a first step towards the solution. A numerical and asymptotic analysis of
this type of models can be found in \cite{Bobev:2014jva}\footnote{Cf.~also \cite{Suh:2018vbp}.},
where a particular truncation of $N=8$, $d=5$ gauged supergravity is studied.

\section{Conclusions}
\label{sec:concl}

In this paper, we used the residual symmetries of $N=2$, $d=4$ Fayet-Iliopoulos-gauged supergravity 
discovered in \cite{Cacciatori:2016xly} to add an electric charge density and rotation to five-dimensional
black strings that asymptote to AdS$_5$. This is the first instance of using solution-generating techniques
in gauged supergravity to add rotation to a given seed, and opens the possibility to construct in a
similar way many solutions hitherto unknown, which are potentially interesting particularly in an
AdS/CFT context.

The rotating string \eqref{eq:metrbs2} interpolates between magnetic AdS$_5$
at infinity and a deformation of $\text{AdS}_3\times\text{H}^2$ near the horizon. This deformation
implies that the CFT$_2$, to which the dual four-dimensional CFT flows in the IR, has less symmetry.
We did not check explicitely how many supercharges are preserved by the near-horizon metric
\eqref{nh-rot-string}, but we would expect that the rotation breaks at least some of the original
supersymmetries.

We also constructed static magnetic BPS black strings with running scalars
in the FI-gauged stu model. It was shown that this amounts to solving the $\text{SO}(2,1)$ spinning top 
equations, which descend from an inhomogeneous version of the Nahm equations.
We were able to solve these in a particular case, which leads to a generalization of the
Maldacena-Nu\~nez solution. Moreover, we computed the central charge of the CFT$_2$ dual to
the near-horizon configuration. From the behaviour of the two bulk scalar fields $\phi^1$ and $\phi^2$
near the conformal boundary
we saw that they correspond in the dual SCFT to an insertion and an expectation value of an operator of 
scaling dimension $\Delta=2$.
The relevant deformation of the dual superpotential relative to $\phi^1$ is described
in \cite{Maldacena:2000mw}, while $\phi^2$ is a marginal deformation of two-dimensional
$N=(4,4)$ SYM theory. Thus, our solution does not describe the gravity dual of 2d 
$N=(2,2)^*$ SYM \cite{Nian:2017usa}.

We hope to come back in particular to further applications of the duality transformations of
\cite{Cacciatori:2016xly} in the near future.

\appendix

\section*{\label{appendix}Appendix}

\section{Supersymmetry variations}
\label{sec-SV}

The supersymmetry variations of the gravitino $\psi_\mu$ and the gauginos $\lambda_i$ in
$N=2$, $d=5$ FI-gauged supergravity coupled to vector multiplets
read \cite{Gunaydin:1984ak}

\begin{equation}
\begin{split}
&\delta\psi_{\mu} = \left(D_\mu + \frac i8 h_I ({\Gamma_\mu}^{\nu\rho} - 4{\delta_{\mu}}^{\nu} 
\Gamma^\rho) F^I_{\nu\rho} + \frac12 g\Gamma_\mu h^I V_I\right)\epsilon\,, \\
&\delta\lambda_i = \left(\frac38\Gamma^{\mu\nu} F^I_{\mu\nu}\partial_i h_I - \frac i2
\mathcal G_{ij}\Gamma^\mu\partial_\mu\phi^j + \frac{3i}2 g V_I \partial_i h^I\right)\epsilon\,,
\label{eq:BPS}
\end{split}
\end{equation}
where 
\begin{equation}
D_\mu\epsilon = \left(\partial_\mu + \frac14\omega_\mu^{ab}\Gamma_{ab} - \frac{3i}2 g V_I A^I_\mu 
\right)\epsilon\,.
\end{equation}

\section*{Acknowledgements}

This work was supported partly by INFN. We would like to thank A.~Amariti and N.~Petri
for useful discussions and C.~Toldo and A.~Tomasiello for valuable comments.

\end{document}